\documentclass[12pt]{article}

\begin{document}

\begin{center}
{\large {\bf {\Large Electrovac $pp$-waves}}} \\[2mm]  {\bf {\large
Y. Nutku } } \\ Feza G\"ursey Institute
P.O.Box 6 \c{C}engelk\"oy, Istanbul 81220 Turkey \\[4mm]
%deli.tex   \qquad \today
\end{center}

\noindent{\bf Abstract} New exact solutions of the
Einstein-Maxwell field equations that describe $pp$-waves are
presented.

\section{Historical Introduction}

In 1926 it was shown by Baldwin and Jeffrey  \cite{1926} that
\begin{equation}
d s^2 = 2 d u \, d v - 2 d z \, d \bar z + h^2  | z |^2 \, d u^2,
\label{1}
\end{equation}
with $h=h(u)$ an arbitrary function, is an exact solution of the
Einstein-Maxwell field equations. We have here taken only the
germane part of the Baldwin-Jeffrey solution. The Maxwell field is
given by
\begin{equation}
F_{u z} = F_{u \bar z}= h \label{bjem}
\end{equation}
which has no dependence on the spatial coordinates. Since then
this solution has been rediscovered several times and is known
under different names.

In particular for $h=1$ it is often called the Bertotti \cite{b}
Robinson \cite{r} solution
\begin{equation}
d s^2 =  \frac{1}{r^2} \left[ d t^2 - d r^2  - r^2 ( d \theta^2 +
\sin \theta \,  d \phi^2 ) \right] \label{br}
\end{equation}
which has the topology $AdS_2 \times S^2$. This solution has most
recently reappeared in string field theory in the form
\begin{equation}
d s^2 = 2 d u \, d v  + a^+ d a^+ + a^- d a^- + ( a^+ d a^- - a^-
d a^+ ) \, d u + b \, d u^2
 \label{nw}
\end{equation}
by Nappi and Witten \cite{nw}.

For $h=\theta(u)$ where $\theta$ denotes the Heaviside unit step
function, the Baldwin-Jeffrey solution is expressible as
\begin{equation}
d s^2 = 2 d u \, d v - \sin^2 [u \theta(u)] \, d x^2 - \cos^2 [u
\theta(u)] \, d y^2
 \label{bs}
\end{equation}
which plays an important role as initial data \cite{n} in the Bell
and Szekeres \cite{bs} Cauchy problem for colliding shock
electrovac waves.

All these metrics are isometric to particular cases of the
Baldwin-Jeffrey solution (\ref{1}), (\ref{bjem}).

\section{Single electrovac $pp$-wave}

In the case of vacuum it has long been recognized that plane waves
are only a particular case of $pp$-waves \cite{ek}. But unlike the
case for vacuum, the $pp$-type generalization of (\ref{1}) to
Einstein-Maxwell fields has not been studied as extensively. The
natural extension of (\ref{1}) is
\begin{equation}
d s^2 = 2\, d u \, d v - 2 \, d z \, d \bar z + h^2 \, | z |^{ 2
n} \, d u^2 \label{newm}
\end{equation}
which is an exact solution of the Einstein-Maxwell field equations
with the Maxwell potential $1$-form
\begin{equation}
A = h \,  ( z^n + \bar z^n ) \, d u \label{newpot}
\end{equation}
and $h$ is again an arbitrary function of $u$. This has been
obtained before \cite{4lu}, \cite{o}.  Note that here $n$ can be
any real number, not necessarily an integer. For $n=1$ we have the
Baldwin-Jeffrey solution (\ref{1}).

\section{Superposition Principle}

In vacuum $pp$-waves travelling in the same direction can be
superposed rather simply. For a collection of electrovac
$pp$-waves of the type (\ref{newm}), (\ref{newpot}) we can also
construct a superposition principle but it is non-trivial. The
space-time metric for the superposition of two electrovac
$pp$-waves is given by
\begin{equation}
d s^2 = 2\, d u \, d v - 2 d z \, d \bar z + \left[ h_1^2 \; | z
|^{ 2 n} + h_2^2 \; | z |^{ 2 m} + h_1 h_2 \left( z^n \bar{z}^m +
\bar{z}^n z^m  \right) \right] \, d u^2 \label{newmet2}
\end{equation}
with the Maxwell potential $1$-form
\begin{equation}
A = \left[ h_1 ( z^n + \bar z^n ) + h_2 ( z^m + \bar z^m ) \right]
\, d u \label{newpot2}
\end{equation}
where $h_1, h_2$ are  arbitrary functions of $u$ and $n, m$ are
arbitrary real numbers. For the Maxwell field we have simple
superposition, but in the metric there is an interference term.
This type of superposition can be carried out for many such fields
in a straight-forward way.

\section{General Solution}

The fact that we can superpose electrovac $pp$-waves suggests that
we should consider a a holomorphic function which is given by a
Laurent series expansion
\begin{equation}
f(u,z)= \Sigma_{i=-\infty}^\infty h_i \, z^i \label{arb}
\end{equation}
where each $h_i$ is an arbitrary function of $u$. Then the Maxwell
potential $1$-form will be
\begin{equation}
A = \left[ f(u,z) + \bar f(u, \bar z ) \right] \, d u,
\label{newpotn}
\end{equation}
which follows from simple superposition (\ref{newpot2}). The
metric that has as source the Maxwell energy-momentum tensor
constructed from (\ref{newpotn}) is given by
\begin{equation}
d s^2 = 2\, d u \, d v - 2\, d z \, d \bar z +  | f(u,z) |^{ 2} \,
d u^2 \label{newmetn}
\end{equation}
where $f(u,z)$ is the arbitrary holomorphic function (\ref{arb}).

\section{Acknowledgement}
I thank Y. Obukhov and M. Ortaggio for interesting and informative
emails on the original version of gr-qc/0502049. I also thank C.
Deliduman for showing me ref. \cite{nw}

\end{document}